\documentclass[preprint,10pt]{elsarticle}
\pdfoutput=1
\usepackage{amssymb}
\usepackage{hyperref} % для вставки гиперссылок
\usepackage{amsmath} % всякая полезная математика от AMS - American Math Society
\usepackage{amsfonts}
\usepackage{graphicx} % для картинок
\usepackage{wrapfig}
\usepackage{ulem}

\usepackage[utf8]{inputenc}

\journal{Astronomy Letters}

\begin{document}

\begin{frontmatter}



\title{Cosmic Gamma-Ray Bursts Detected in the RELEC Experiment Onboard the Vernov Satellite}


\author[1]{A.V. Bogomolov}
\author[1,2]{V.V. Bogomolov}
\author[1]{A.F. Iyudin}
\author[1,2]{E.A. Kuznetsova}
\author[3]{P.Yu. Minaev}
\author[1,2]{M.I. Panasyuk}
\author[3,4]{A.S. Pozanenko}
\author[1,2]{A.V. Prokhorov}
\author[1,2]{S.I. Svertilov\corref{cor1}}
\ead{sis@coronas.ru}
\author[3]{A.M. Chernenko}

\cortext[cor1]{Corresponding author}

\address[1]{
	Skobeltsyn Institute of Nuclear Physics, Lomonosov Moscow State University, Moscow, 119991, Russia}
\address[2]{
	Lomonosov Moscow State University, Moscow, 119991, Russia}
\address[3]{
	Space Research Institute, Russian Academy of Sciences, Profsoyuznaya ul. 84/32, Moscow, 117997, Russia}
\address[4]{MEPhI National Research Nuclear University, Kashirskoe sh. 31, Moscow, 115409, Russia}

\sloppy
\begin{abstract}
The RELEC scientific instrumentation onboard the Vernov spacecraft launched on July 8, 2014, included the DRGE gamma-ray and electron spectrometer. This instrument incorporates a set of scintillation phoswich detectors, including four identical X-ray and gamma-ray detectors in the energy range from 10 keV to 3 MeV with a total area of $\sim$500 $cm^{2}$ directed toward the nadir, and an electron spectrometer containing three mutually orthogonal detector units with a geometry factor of $\sim$2 $cm^{2} sr$, which is also sensitive to X-rays and gamma-rays. The goal of the space ex- periment with the DRGE instrument was to investigate phenomena with fast temporal variability, in particular, terrestrial gamma-ray flashes (TGFs) and magnetospheric electron precipitations. However, the detectors of the DRGE instrument could record cosmic gamma-ray bursts (GRBs) and allowed one not only to perform a detailed analysis of the gamma-ray variability but also to compare the time profiles with the measurements made by other instruments of the RELEC scientific instrumentation (the detectors of optical and ultraviolet flashes, the radio-frequency and low-frequency analyzers of electromagnetic field parameters). We present the results of our observations of cosmic GRB 141011A and GRB 141104A, compare the parameters obtained in the GBM/Fermi and KONUSWind experiments, and estimate the redshifts and Eiso for the sources of these GRBs. The detectability of GRBs and good agreement between the independent estimates of their parameters obtained in various experiments are an important factor of the successful operation of similar detectors onboard the Lomonosov spacecraft.

\end{abstract}




\end{frontmatter}

\sloppypar 
\vspace{2mm}
\section*{INTODUCTION}
\label{}
\noindent
Cosmic gamma-ray bursts (GRBs) are known to be among the most powerful phenomena in the Universe. Two groups of events are identified among them: with a duration of more than 2 s (long GRBs) and less than 2 s (short GRBs). The latter are characterized by, on average, harder gamma- ray spectra than those for long bursts. The long bursts are associated with the core collapse of a star with a mass of tens of solar masses (the hypernova model; see, e.g., Paczynski 1998), while the short ones are associated with the merger of compact relativistic objects in a binary system (a neutron star with a neutron star or a neutron star with a black hole; see, e.g., Woosley 1993). In both cases, the peculiarities of the explosive process during the collapse (central engine operation) are such that energy is released asymmetrically and relativistic jets outflowing from the poles of a rapidly rotating collapsing body are formed. Despite the abundance of theoretical models, the details of the central engine operation and the jet formation mechanisms have not yet been completely clarified. In particular, one of the fundamental problems in the physics of cosmic GRBs is the necessity of explaining the temporal burst structure.

A detailed analysis of such temporal structures requires a high time resolution of detectors. This is particularly topical in investigating short GRBs. The best conditions for investigating the fast variability of GRBs are achieved when recording individual photons (the so-called photon-by- photon record). This mode was envisaged in the gamma-ray detectors of the DRGE instrument included in the RELEC instrumentation onboard the Vernov satellite (Panasyuk et al. 2016). Although the main goal of this experiment was to study terrestrial gamma-ray flashes (TGFs) (Bogomolov et al. 2017), it was also possible to detect cosmic GRBs during its operation. For example, two cosmic GRBs were detected during the observations onboard the Vernov satellite from July to December 2014, which were confirmed by the data of other spacecraft. The short GRB 141011A with a duration of 75 ms, whose time profile exhibits a fine temporal structure, is of special interest among them. This paper is devoted to the investigation of the GRBs detected by the DRGE instrument and to the comparison with other experiments. The main parameters of the DRGE experiment are presented in Section 1. The data for GRB 141011A and GRB 141104A are analyzed in Sections 2-4. The results obtained are discussed in Section 5.

%*************************************************************
\section{EXPERIMENTAL CONDITIONS AND OBSERVING TECHNIQUE}
\noindent
The RELEC experiment was realized onboard a small (MKA-FKI PN2) spacecraft named Vernov in honor of Academician Sergei Nikolaevich Vernov, the founder of Russian space physical science. The spacecraft is based on the Karat platform designed and manufactured by the S.A. Lavochkin Research and Production Association (Khartov et al. 2011). The satellite was launched on July 8, 2014, and had the following characteristics:

mass --- 283 kg;

orientation accuracy --- 6 arcmin;

stabilization accuracy  --- 0.0015$^{\circ}s^{-1}$;

data transfer rate --- 5 Mbit $s^{-1}$. 

operational satellite orbit — solar-synchronous with an apogee of 830 km, a perigee of 640 km, an inclination of 98.4$^{\circ}$, and a revolution period of 100 min.

The scientific instrumentation provided the monitoring measurements of high-energy electron fluxes with the possibility of determining the flux anisotropy, the detection of atmospheric transient events in a wide range of the electromagnetic spectrum (from radio to gamma-rays) with a high time resolution, and the measurements of electric and magnetic field pulsations in the frequency range from 0.1 Hz to 15 MHz.

The GRBs were detected with the DRGE-1 and DRGE-2 units of the DRGE instrument, which recorded the fluxes and energy spectra of hard electromagnetic radiation with a gamma-ray energy in the range $E=0.01-3.0$ MeV, electrons with$E=0.02-15$ MeV, and protons with $E=4-100$ MeV (Panasyuk et al. 2016).
 
\begin{table}[t]
	
	\vspace{6mm}
	\centering
	\caption {Physical and technical characteristics of the
		DRGE instrument onboard the Vernov satellite}\label{DRGE} 
	
	\vspace{5mm}\begin{tabular}{l|c} 
		\hline
		\hline
		{Energy range}&$kT,$ \\ 
		photons  & 0.01--3 MeV \\
		electrons  & 0.5--10 MeV \\
		protons  & 10--100 MeV \\
		\hline
		Effective area  & $4\times120=480$ cm$^2$ \\
		of scintillation detector   & (for 4 detectors) \\
		\hline
		Field of view  & $2\pi$ sr ($\pm90^{\circ}$) \\
		\hline
		Mass   & $\sim$10.4 kg (for each unit) \\
		\hline
		Sizes   & ($0.36\times0.36\times0.18$) m$^3$ \\
		\hline
		Data volume   & $\sim$150 MB$/$day \\
		\hline
		Power consumption   & $\sim$9 W (for each unit) \\
		\hline
	\end{tabular}
\end{table}
The DRGE instrument comprised three units: two identical DRGE-1 and DRGE-2 units and one DRGE-3 unit. The physical and technical characteristics of the instrument are given in Table \ref{DRGE}. Each of the DRGE-1(2) units consisted of two identical detector units (DRGE-11, DRGE-12 and DRGE-21, DRGE-22) including a scintillation phoswich detector and a photomultiplier tube (PMT). The scintillator of the phoswich detector is made of NaI(Tl) crystals with a thickness of 0.3 cm and a diameter of 13 cm and CsI(Tl) with a thickness of 1.7 cm and a diameter of 13 cm. Both detectors were in optical contact and were placed in a protective cover. The detectors of the DRGE-1 and DRGE-2 units were oriented toward the nadir. To a first approximation, the dependence of the effective area of the detector units on the angle of incidence of the emission $\alpha$ can be assumed to be proportional to $\cos(\alpha)$.

During the experiment the count rates of the DRGE-1(2) detectors were continuously recorded in two modes: the monitoring and time-tagged event (TTE) ones. The initialization of an internal clock at the instant the synchronizing pulse arrived from the satellite was used in each unit to ensure the synchronous operation of all detector units with a good time resolution. The clock of each unit had a period of $15.48 \mu s$. The clock stability $\sim10^{-5}$ provided a synchronization accuracy of $\sim 15 \mu s$.

The mean integrated gamma-ray photon count rate separately for each of the NaI(Tl) and CsI(Tl) crystals and the total count rate in both crystals were measured in the monitoring mode for each detector unit in the exposure time (1 s). The minimum threshold energy of a recorded photon was $\sim$10 keV for NaI(Tl) and $\sim$25 keV for CsI(Tl). Thus, in the monitoring mode there were almost continuous time series of 1-s-averaged count rates of gamma-ray photons with energies $>10$ keV in NaI(Tl) and $>25$ keV in CsI(Tl).

In the TTE mode the detection time and amplitude codes were recorded for each detected gamma-ray photon, which allowed one to determine the part of the detector where the interaction occurred and what energy was released in the crystals. In view of the existing constraints on the volume of transmitted information, constraints on the number of successively recorded events were imposed. No more than 800 events were successively recorded in each second for each detector in the case of a relatively low background count rate ($<$1000 counts $s^{-1}$ in both crystals). If the background count rate was >1000 and >1500 counts $s^{-1}$, then no more than 200 and 50 events, respectively, were recorded. Near the equator the total background in both crystals was less than 800 counts $s^{-1}$, i.e., the data on all the detected gamma-ray photons were recorded. Near the polar caps the total background was $>1000$ counts $s^{-1}$), i.e., only about a fifth of the events detected per second were recorded.

The data obtained in the TTE mode were mainly used to find the candidates for TGFs, because they are characterized by relatively short durations ($<$1 ms). The condition for a significant excess above the mean background count rate in a time interval of 1 ms simultaneously in at least two detectors was used in this case. Since TGFs are characterized by a hard spectrum, we selected only the events that gave a total energy release $>400$ keV in both crystals. The mean background count rate of such events on the equator was no more than one event in several milliseconds, and the requirement for the detection of no fewer than five gamma-ray photons in 1 ms by at least two detectors or no fewer than three gamma-ray photons by at least three detectors was chosen as a significance criterion. For the equator this actually corresponds to the simultaneous detection of increases by two detectors at a 12$\sigma$ confidence level or by three detectors at a 7$\sigma$ confidence level. In this case, not the number of counts in the above time interval but the duration of the intervaloccupied by several successively recorded events was determined. Thus, the following requirement corresponds to this criterion: for five successively recorded events the duration of the interval between the first and the last should not exceed 1 ms. Furthermore, an additional condition was used to exclude the detection of increases due to intense background variations in the regions of trapped radiation: the mean background count rate in an interval of $\pm$1 s from the increase being analyzed should not exceed 1500 counts $s^{-1}$.

\begin{figure}
	%\vspace{6cm}
		\centering
	\includegraphics[width=10cm]{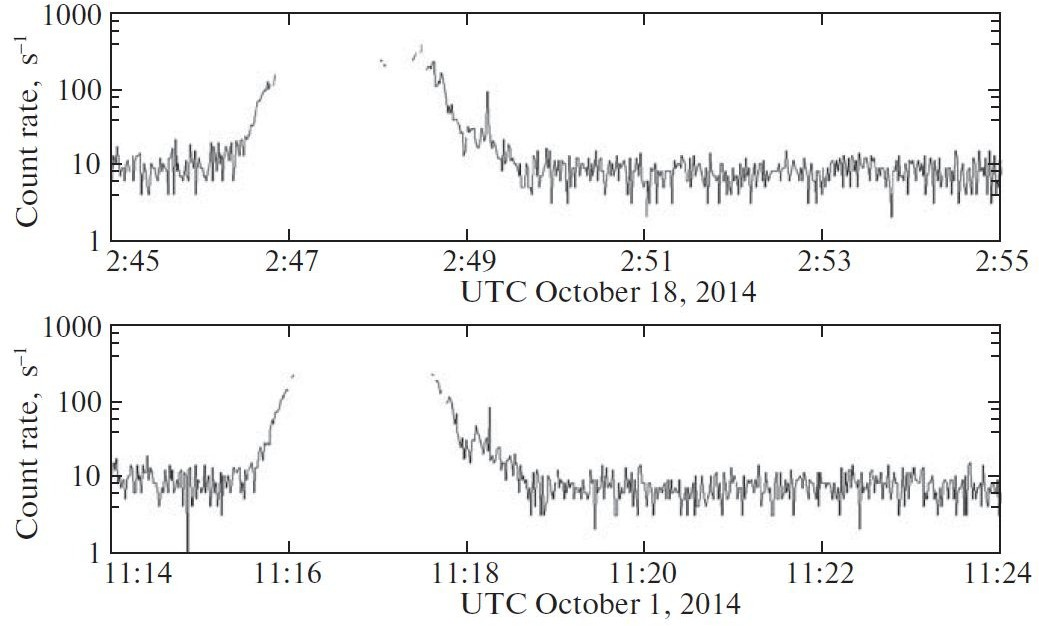} 
	\caption{The DRGE 31 count rates illustrating the possibility of a GRB imitation by electron 
uxes. (a) The measurements during the October 18, 2014 event that triggered KONUS-Wind (approximate time 2 : 49, L = 6:73). (b) The measurements on October 1, 2014, for a similar trajectory of the Vernov satellite (approximate time 11:18, L = 6:35).} 
	\label{fig1}
\end{figure}

By applying the criterion considered above, we selected the candidates for TGFs (Bogomolov et al. 2017); we developed a special technique for eliminating the imitations of a short GRB by heavy charged Galactic cosmic-ray particles. Among the bursts that satisfied the above criteria there was one detected on October 11, 2014, at 06:46:20 UT with a duration of $\sim$50 ms, which exceeds considerably the duration of a typical TGF. By comparison with the data from the Gamma Ray Coordinates Network (GCN) [http://gcn.gsfc.nasa.gov], this burst was identified as a cosmic one, which was also observed at least in six space experiments: Fermi (GBM) (von Kienlin 2014), Konus-Wind (Golenetskii et al. 2014b), MESSENGER (GRNS), Mars Odyssey (HEND), Swift (BAT), and INTEGRAL (SPI-ACS), and was initially localized by the Fermi (GBM) experiment with an accuracy of 2.7 deg (von Kienlin 2014), and the localization accuracy was then improved by triangulation (Golenetskii et al. 2014a) using the data of above space experiments of the IPN network (\verb$http://ssl.berkeley.edu/ipn3/$).

In contrast, the other cosmic GRB was identified using the data obtained in the monitoring mode, when the bursts from GCN notices were checked for the corresponding response in the DRGE-1(2) detector count rates. This GRB was detected on November 4, 2014, at 00 : 03 : 21 UT; it was also observed in five space experiments: Swift (BAT), Konus-Wind (Golenetskii et al. 2014c), MESSENGER (GRNS), Mars Odyssey (HEND), and INTEGRAL (SPI-ACS), and was
also localized by triangulation (Hurley et al. 2014). Below these bursts will be analyzed in detail. During much of the time the Vernov satellite was in the Earths radiation belts due to the peculiarities of its orbit. Investigating the detectability of GRBs in such orbits is important for the next experiment aimed at detecting bursts onboard the Lomonosov satellite, which has a similar orbit. In such orbits the increases in count rates in gamma-ray detectors resembling a GRB in intensity and duration can also be caused by magnetospheric electrons. Such phenomena are observed quite often, and some of them can by chance coincide in time with a real GRB detected in other experiments. As an example of such an event, Fig. \ref{fig1} presents the count rates of the DRGE- 31 unit. A flux increase coincident in time with the trigger of GRB 141018 in the Konus-Wind experiment at $\sim$2 : 49 : 12 UT (\verb|http://gcn.gsfc.nasa.gov/konus 2014grbs.html|) is observed on the upper panel. However, a similar increase detected in the same place at a different time (see the lower panel in Fig. \ref{fig1}) suggests that this increase is not a cosmic GRB. Such increases are most commonly observed near the Earths outer radiation belt from the polar caps at L $\sim6 - 7$ and higher or from the equator. In total the Vernov satellite spent about 45\% of the time in the (outer and inner) radiation belts with a high background level.
Both GRBs mentioned above occurred when the Vernov satellite was near the equator, i.e., under favorable background conditions. For both GRB 141011A occurred at L $\sim$1.15 and GRB 141104A occurred at L $\sim$1.7 the background count rates were approximately constant during a time exceeding considerably the burst duration. Therefore, it turned out to be possible to determine the background level empirically by averaging the count rates within several tens of seconds before and after the burst in the same energy ranges as those for the bursts. The burst spectra were fitted by various models after the subtraction of the background values from the count rates in each individual range.

\section{GRB 141011A}
\noindent
The source of this burst was located in such a way that the gamma-ray flux was incident at a very large angle to the zenith-nadir line of the DRGE- 1(2) detectors and, therefore, exposed the entrance windows of these detectors oriented toward the nadir. The angle with the direction toward the nadir was 77$^{\circ}$. To determine this angle, we used the coordinates of the center of the sources error region, $R.A.=$17 h 11 m 45 s and $Dec.=-9^{\circ} 40^{\prime} 52^{\prime\prime}$, obtained by triangulation from the data of several spacecraft (Golenetskii et al. 2014a).

\begin{figure}
	%\vspace{6cm}
	\centering
	\includegraphics[width=8cm]{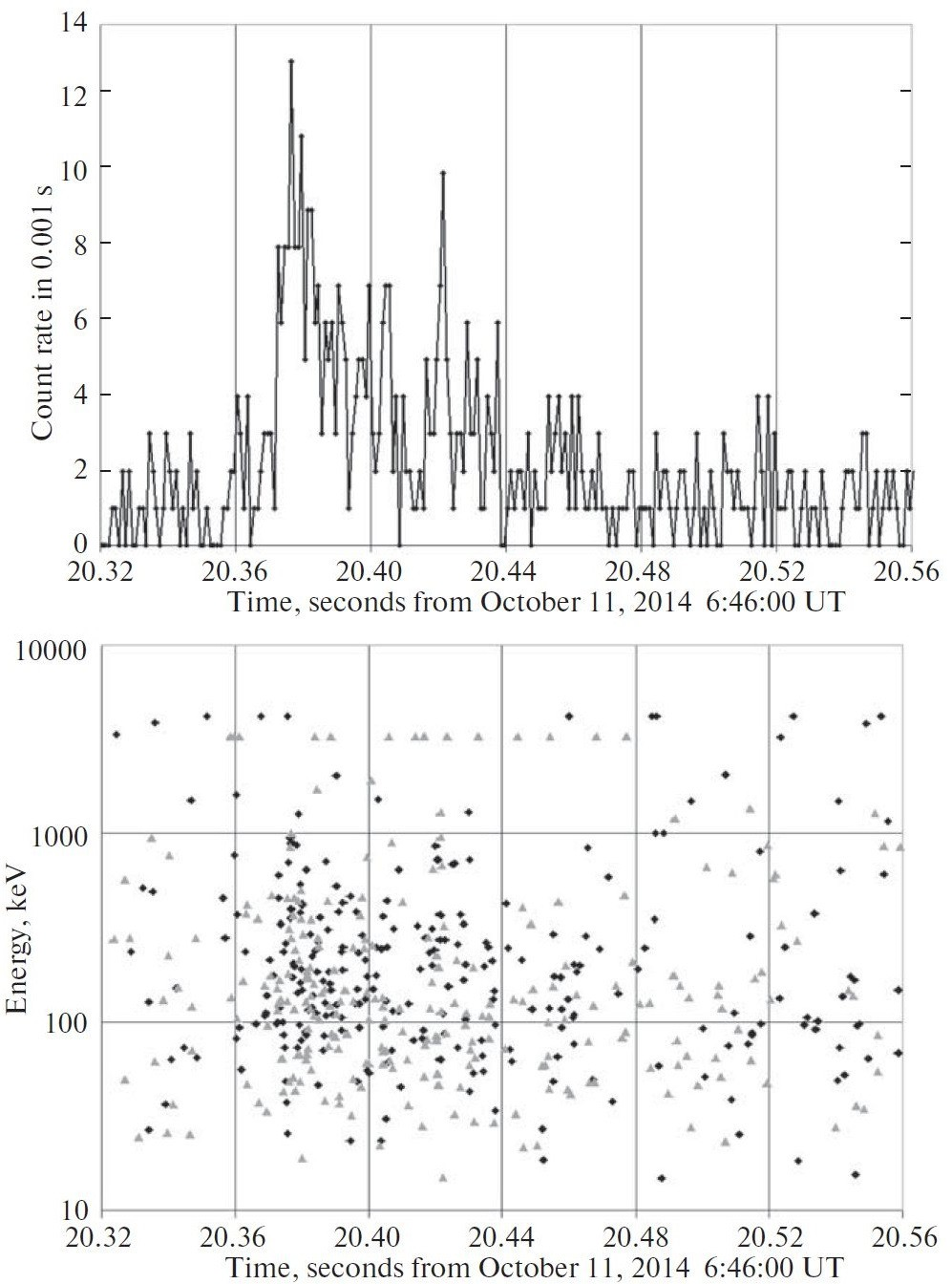} 
	\caption{(a) Light curve of the short GRB 141011A based on the Vernov satellite data from 6:46 UT (the sum of the count rates from the DRGE-11 and DRGE-12 detectors of the RELEC instrumentation) with a time resolution of 0.001 s. (b) The energytime diagram obtained during GRB 141011A by the DRGE-11 (black diamonds) and DRGE-12 (gray triangles) detectors onboard the Vernov satellite. Each point corresponds to one photon.} 
	\label{fig2}
\end{figure}

\begin{figure}
	%\vspace{6cm}
	\centering
	\includegraphics[width=8cm]{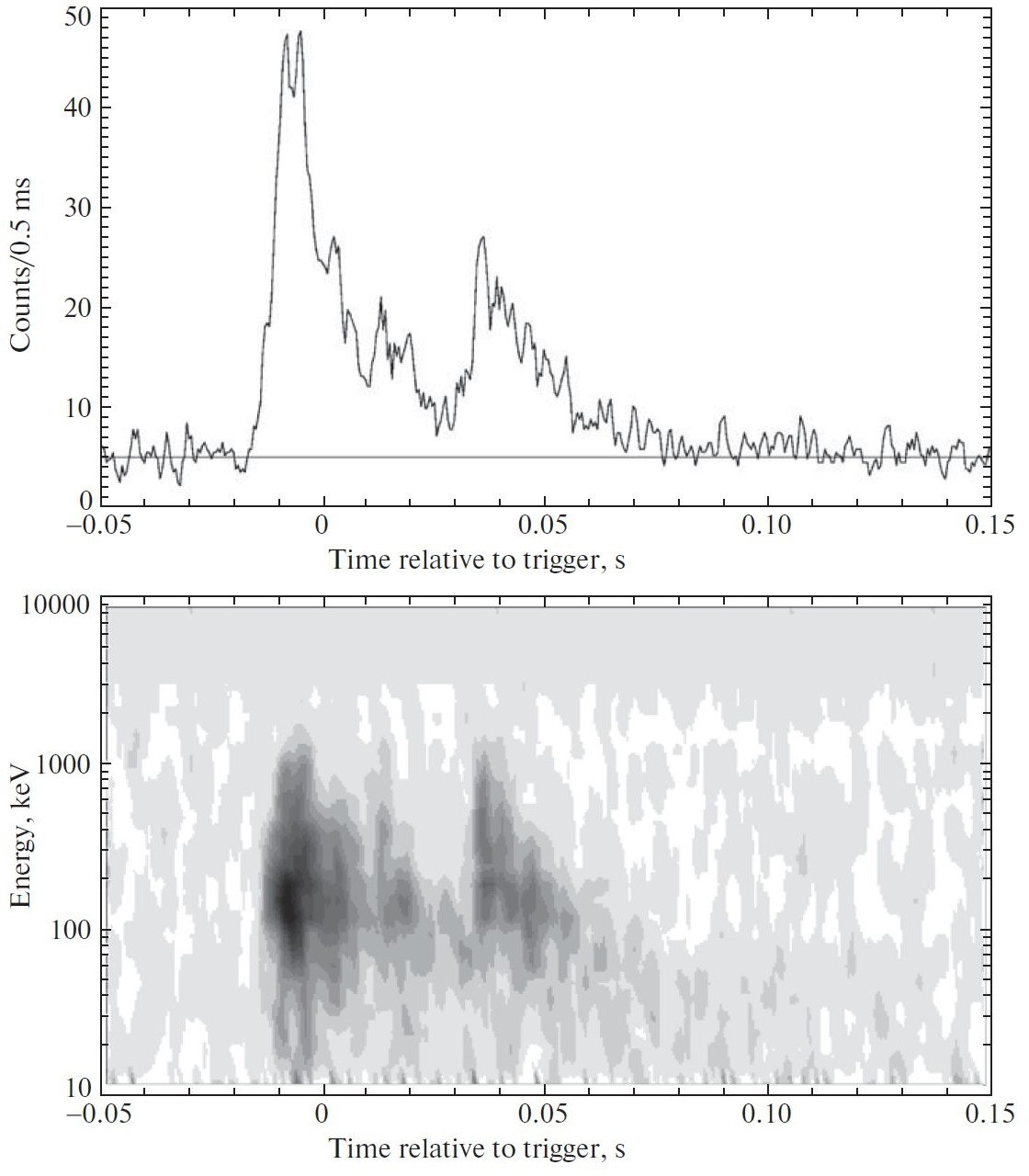} 
	\caption{GBM/Fermi (NaI: 0--3, 9, 10; BGO: 0, 1) light curve of the short GRB 141011A: (a) the light curve with a time resolution of 0.5 ms in the energy range 10--10 000 keV and (b) the energy diagram. The background level is shown in the light curve. A darker shade on the diagram corresponds to a higher 
ux. The spectral evolution from hard to soft radiation is seen.} 
	\label{fig3}
\end{figure}

The GRB time profiles with a time resolution of 1 and 0.5 ms constructed using data from the DRGE- 1(2) detectors of the RELEC instrumentation onboard the Vernov satellite and the GBM/Fermi instrument, respectively, are presented on the upper panels of Figs. \ref{fig2} and \ref{fig3}. A fine temporal structure in the form of two separate intense peaks spaced $\sim$50 ms apart ($\sim$6 : 43 : 20.38 and $\sim$6 : 43 : 20.42 s) is traceable in the time profiles. The front of themain peaks is milliseconds, while the decay is considerably slower, being tens of milliseconds. Several weaker peaks ($\sim$6 : 43 :
20.40 and $\sim$6 : 43 : 20.45 s) are observed at the decay of the main peaks.
The temporal structure of the burst can be traced in more detail, on even shorter time scales, on the energy-time diagrams shown on the lower panels of Figs. \ref{fig2} and \ref{fig3}. The detection time of each event is along the horizontal axis on these diagrams; the total energy release in both crystals of the detector unit under consideration is along the vertical axis. Thus, each point on the diagram reflects the recorded event, while the diagram as a whole reflects the energy distribution of events as a function of time. In the RELEC experiment the noise track in the range of low energy releases corresponds to the signals in the PMT and the scintillators due to the fluctuations in the number of thermal photons. When recording the events that give an energy release above the limit of the dynamic range, in particular, when cosmic-ray particle hit the detector, a track corresponding to maximum energy releases and associated with electronics overloads emerges. This explains the presence of points with energy releases of $\sim$3 MeV (DRGE-11) and $\sim$4 MeV (DRGE-12) on the lower panel of Fig. \ref{fig2}, which are always present both during the burst and at a different time. A significant cluster of points is observed during the maxima of the GRB light curve ($\sim$0.38 and
$\sim$0.42 in Fig. \ref{fig2}).

A considerable increase in the number of photons with an energy above 500 keV for several mil- liseconds exclusively at the maxima of the peaks, which was not observed between the maxima, in contrast to lower-energy photons, should be particularly noted. For greater clarity, when analyzing the fine temporal structure at a high intensity of radiation, Fig. \ref{fig2} shows the data only from one of the units, DRGE-11. The availability of complete information about the energy and detection time of each photon allows the light curves to be constructed in any energy subband and with a time resolution whose expediency is determined only by the statistics (for GRB 141011A we chose an averaging time of 1 ms).

To obtain the spectral GRB characteristics, we modeled the DRGE-1(2) detector response to gamma-rays incident at various angles to the detector axis. We used three main spectral models: 
\begin{enumerate}
\item	a simple power law (PL):  
\begin{equation*}
N(E)=AE^{\alpha}
\end{equation*}
\item a power law with an exponential cutoff (CPL): 
\begin{equation*}
N(E)=AE^{\alpha}\exp\left( \frac{-E(2+\alpha)}{E_p} \right)
\end{equation*}

\item	a power law with a break (BAND, Band et al. 1993):
\begin{equation*}
N(E) = 
\begin{cases}
A\left( \dfrac{E}{100} \right) ^{\alpha}\exp\left( \dfrac{-E(2+\alpha)}{E_p} \right),   & E\leq\dfrac{(\alpha-\beta)E_p}{2+\alpha}\\
A\left( \dfrac{(\alpha-\beta)E_p}{100(2+\alpha)} \right) ^{\alpha-\beta}\exp(\beta-\alpha)\left( \dfrac{E}{100} \right) ^{\beta},   & E>\dfrac{(\alpha-\beta)E_p}{2+\alpha}
\end{cases}
\end{equation*}
\end{enumerate}

The function proposed by Band is a superposition of a power law with an exponential cutoff (the exponent $\alpha$ is used in the formula), which, as a rule, describes well the spectra of bursts at low energies, and a power law (the exponent $\beta$ is used in the formula) for high energies. The coefficients in the formula were chosen in such a way that the function remained continuous when passing from the upper part to the lower one. The parameter $E_p$ is the so-called peak energy that corresponds to the maximum in the spectral representation $N(E)E^{2}$ (Band et al. 1993; Ford et al. 1995).
\begin{figure}
	%\vspace{6cm}
	\centering
	\includegraphics[width=10cm]{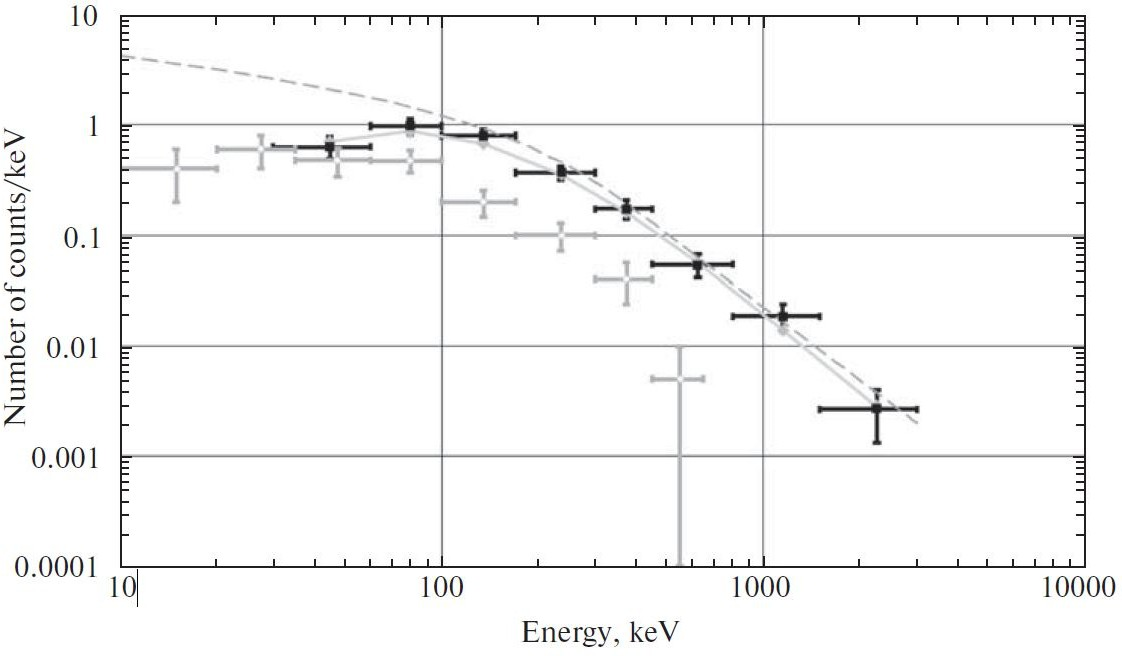} 
	\caption{Energy spectrum of GRB 141011A based on the Vernov satellite data. The intervals indicate the experimental data: the count rate in Cs I(Tl) (black intervals), the count rate in NaI(Tl) (gray intervals). The dashed line is the GRB spectrum (BAND model) reconstructed from the count rates in Cs I(Tl). The solid line is a convolution of the model function with the instrument response matrix based on the Cs I(Tl) data.} 
	\label{fig4}
	\end{figure}
For each spectral model and combination of parameters we calculated the expected response of the detector oriented at a specified angle of 77$^{\circ}$ to the source in eight NaI(Tl) intervals and eight CsI(Tl) intervals coincident in energy with the monitoring intervals. The calculations were performed using the MEGAlib toolkit v.2.29.01 software package based on GEANT4 (geant4-10-00-patch-03) (Zoglauer 2005; Zoglauer et al. 2006). Using the goodness-of-fit $\chi^2$ test, we then compared the experimental data with the model ones separately for each fixed combination of parameters ($\alpha$ for PL, $E_p$ and $\alpha$ for CPL, and $E_p$, $\alpha$, and $\beta$ for BAND). In each case, we first normalized the model data: the coefficient was established in such a way that the sum of the model count rates over all the investigated energy intervals was equal to the sum of the measured count rates in the same intervals. Thereafter, we calculated the $\chi^2$ value and the significance level of the statistical hypothesis about the equality of the experimental and model values. The number of degrees of freedom in this case is smaller by one than the number of intervals. No additional decrease in the number of degrees of freedom due to the presence of parameters $E_p$, $\alpha$, and $\beta$ occurs, because they are prespecified rather than determined from the experimental data. At the final stage, once $\chi^2$ and the significance levels have been obtained for all combinations of model parameters, we chose the one for which the $\chi^2$ value was minimal, while the significance level was maximal.

The burst energy spectrum constructed over the entire interval (from 6 : 46 : 20.37 to 6 : 46 : 20.44 UT) is shown in Fig. \ref{fig4}, which also presents its fit by the BAND model. We obtained the following parameters of this model: $E_p= 500\pm150 \mbox{keV, } \alpha =-0.35\pm0.15, \beta=-2.3\pm0.3$. In addition to these parameters, Table \ref{GRB} also gives the values obtained when fitting this spectrum by other models. The corresponding significance levels, along with the $\chi^2$ values and the degrees of freedom used to obtain them, are also given in Table \ref{GRB}. The BAND and CPL models are seen to fit well the spectrum (at a significance level $>0.995$). The power law describes the energy spectrum for GRB 141011A much more poorly.

 \begin{table}[t]
 	
 	\vspace{6mm}
 	\centering
 	\caption{ Parameters of GRB 141011 and GRB 141104 based on the Vernov satellite data.}\label{GRB} 
 	
 	\vspace{5mm}\begin{tabular}{c|c|c|c|c|c|c} 
 		\hline
 		\hline
 		{Event}&Model&$\alpha$&$\beta$&$E_p$, keV&Fluence in range&Significance \\ 
 		&  &  &  &  &2--3000 keV,&level \\ 
 		&  &  &  &  &$10^{-6} erg cm^2$&($\chi^2/dof$)\\
 		\hline
 	{GRB141011} & PL & $-1.3\pm0.2$ & - & - & & 0.931 (2.450$/$7) \\
 		& CPL & $-0.7\pm0.3$ & - & $600\pm200 $& $1.6\pm0.6$ & 0.995 (0.973$/$7)\\
 		& BAND & $-0.35\pm0.15$ & $-2.3\pm0.3$ & $500\pm150$ &  & 0.998 (0.753$/$7)\\
 		\hline
 	{GRB141104} & PL & $-1.1\pm0.2$ & - & - &  & 0.877 (3.082$/$7) \\
 			& CPL & $-0.8\pm0.2$ & - & $800\pm200$ & {$95\pm30$} & 0.975 (1.698$/$7)\\
 		& BAND & $-0.2\pm0.1$ & $-2.4\pm0.4$ & $200\pm60$ &  & 0.969 (1.811$/$7)\\
 		\hline 
 	\end{tabular}
 \end{table}

The lower accuracy (than that in the GBM/Fermi experiment) in determining the best-fit parameters for all models is related primarily to the small effective area of the DRGE instrument for the burst under consideration, especially at low energies, because it was observed at a large angle (77$\circ$) to the detector axis. Nevertheless, satisfactory agreement with the GBM/Fermi data should be noted (see Section 3).

The above estimates of the model parameters were obtained using only the CsI(Tl) channels. Using the softer NaI(Tl) channels is inefficient both because the burst spectrum is hard and because it is difficult to take into account the absorption of the soft radiation incident on the detector at a large angle, which can lead to systematic errors.

\section{INVESTIGATION OF GRB 141011 BASED ON GBM/FERMI DATA}
\noindent
The light curve in the energy range 10 keV – 10 MeV with a time resolution of 0.5 ms for GRB 141011 is presented on the upper panel of Fig. \ref{fig3}. It has a complex shape and consists of two structures with a total duration T90 = 75$\pm$5 ms. In turn, each of the structures consists of several overlapping pulses. This burst is among 10\% of the shortest bursts in the GBM/Fermi experiment (von Kienlin et al. 2014). The light curves of some short bursts are characterized by an additional component, extended emission (see, e.g., Minaev et al. 2010; Svinkin et al. 2016). No extended emission was detected in the light curve of GRB 141011. In both DRGE and GRB/Fermi data we did not detect a burst precursor, an episode before the main phase, either. However, for a sample of short bursts precursors, if they exist (Troja et al. 2010), are detected very rarely (Minaev et al. 2017).

A complex structure of GRB 141011 is also traceable on the energy diagram (the lower part of Fig. \ref{fig3}), which is a two-dimensional distribution of recorded counts in energy-time representation. Darker areas on the diagram correspond to higher fluxes. The spectral evolution from hard to soft radiation for the first and second structures is clearly seen on the diagram.

To estimate the minimum temporal variability scale in the energy range (10---10 000) keV, we constructed the differential light curve, which is the dependence of $D_i=(C_i-C_{i-1})/(C_i+C_{i-1})^{0.5}$ on time $T_i$, where $C_i$ and $C_{i-1}$ are the fluxes of the ordinary light curve at times $T_i$ and $T_{i-1}$, respectively. $D$ reflects the change of the flux in two successive time bins of the light curve expressed in units of the standard deviation. At $T=T_0-0.01$ s relative to the GBM trigger, which corresponds to the phase of a fast rise in flux at the burst onset (see Fig. \ref{fig3}), D is $4.4\sigma$ for the light curve with a time resolution of 2 ms. This means that the minimum statistically significant variability time scale for GRB 141011 is 2 ms.

For the light curves with a time resolution of 0.5 ms in the energy ranges (8--200) and (200--900) keV, we performed a cross-correlation analysis similar to the technique of Band (1997). The maximum of the cross-correlation function is at $lag = 0.0 s$. This means that the spectral lag between these light curves does not exceed 0.5 ms (the time resolution of the investigated light curves). Such a behavior is typical for the class of short GRBs and, on the whole, reflect the tendency for the lag to decrease with decreasing duration of individual pulses (Hakkila and Preece 2011; Minaev et al. 2014). In addition, it can be associated with the superposition effect that arises when investigating a GRB with a complex multipeak structure of the light curve (for more details, see Minaev et al. 2014).
\begin{table}[t]
	
	\vspace{6mm}
	\centering
	\caption{Results of our spectral analysis of GRB 141011 based on the GBM/Fermi data.}\label{spectral} 
	
	\vspace{5mm}\begin{tabular}{l|c|c|c|c|c|c} 
		\hline
		&	&	&	&	&Fluence in range& \\ 
		Time interval, s&Model&$\alpha$&$\beta$&$E_p$, keV&10--1000 keV,&CSTAT/dof\\ 
		&  &  &  &  &$10^7$ erg cm$^{-2}$& \\
		\hline
		(-0.02, 0.07) & PL & $-1.322\pm0.015$ & - & - & $6.99\pm0.19$ & 1106.8$/$791 \\
		& CPL & $-0.502\pm0.075$ & - & $639\pm62$ & $10.54\pm0.38$ & 775.3$/$790\\ 
		& BAND & $-0.491\pm0.085$ & $-2.9^{+0.4}_{-6.6}$ & $618\pm75$ & $10.38\pm0.43$ & 774.5$/$789\\ 
		(-0.02, 0.02) & PL & $-1.293\pm0.018$ & - & - & $4.33\pm0.14$ & 1014.4$/$791 \\
		& CPL & $-0.43\pm0.09$ & - & $704\pm75$ & $7.00\pm0.29$ & 755.4$/$790\\
		& BAND & $-0.43\pm0.09$ & $-4.9^{+1.5}_{-0.6}$ & $712\pm79$ & $7.01\pm0.29$ & 755.3$/$789\\ 
		(0.02, 0.07) & PL & $-1.368\pm0.026$ & - & - & $2.65\pm0.13$ & 833.5$/$791 \\
		& CPL & $-0.56\pm0.013$ & - & $495\pm85$ & $3.51\pm0.25$ & 742.0$/$790\\
		& BAND & $-0.51\pm0.18$ & $-2.4^{+0.4}_{-1.7}$ & $443\pm126$ & $3.38\pm0.25$ & 740.9$/$789\\
		\hline 
	\end{tabular}
\end{table}
We performed a spectral analysis of both the entire GRB (the time interval (-0.02, 0.07) s relativeto the trigger) and its two separate structures (the time intervals (-0.02, 0.02) and (0.02, 0.07) s) using the RMfit v4.3.2 software package (\verb|http://fermi.gsfc.nasa.gov/ssc/data/analysis/rmfit/|). For each time interval we fitted the energy spectrum by three models: a simple power law (PL), a power law with an exponential cutoff (CPL), and a power law with a break (BAND, see Band et al. 1993). The results of our spectral analysis are presented in Table \ref{spectral}. To fit the spectra and to choose the optimal spectral model, we used a modified Cash statistic (CSTAT, see Cash 1979). The simple power-law model describes unsatisfactorily the observed spectrum. The power-law model with an exponential cutoff (CPL) turned out to be the optimal spectral model for all the investigated spectra. The more complex power-law model with a break (BAND model) improves insignificantly the quality of the fit (see Table \ref{spectral}, the values in the last CSTAT/dof column).

\begin{figure} [h]
	%\vspace{6cm}
	\centering
	\includegraphics[width=8cm]{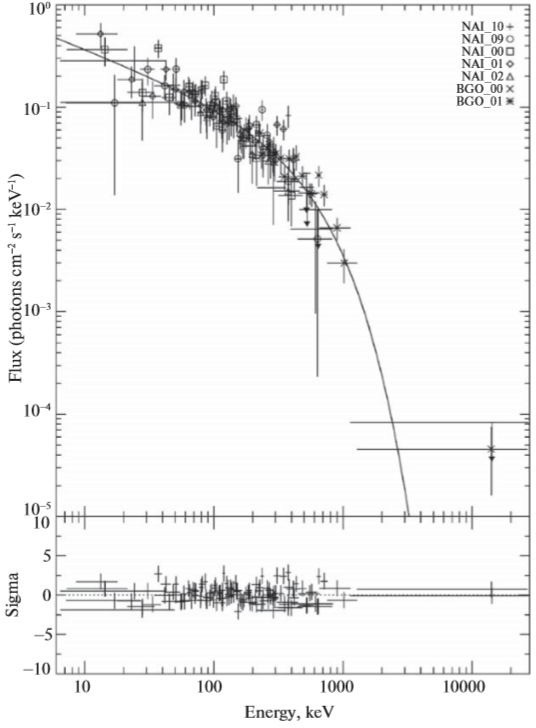} 
	\caption{GBM/Fermi energy spectrum of GRB 141011A in the time interval (-0.02, 0.07) s relative to the trigger. A power-law fit with an exponential cutoff to the spectrum is shown. At the bottom residual is presented expressed in standard deviations.} 
	\label{fig5}
\end{figure}

The energy spectrum of the entire GRB fitted by a power law with an exponential cutoff (CPL) with $\alpha = 0.502\pm0.075\mbox{ and }E_p = 639\pm62$ keV is presented in Fig. \ref{fig5}. The fluence in the energy range 10--1000 keV is $F=(1.05\pm0.04)\cdot10^{-6}$ erg cm$^{-2}$. The spectral hardness calculated as the ratio of the total flux in the range 50--300 keV to the flux in the range 10--50 keV, which are expressed in photons and were calculated in terms of the CPL spectral model, is $HR = 2.0\pm0.2$. On the durationspectral hardness diagram constructed for the GRBs of the GBM/Fermi experiment (Fig. 6 in von Kienlin et al. (2014)), GRB 141011A falls into the region of moderately hard, very short bursts.

\section{GRB 141104A}
\noindent
The energy spectra of the two separate structures are analogous to the spectrum of the entire burst. The spectrum of the first structure is harder than that of the second one ($E_p = 704\pm75$ and $495\pm85$ keV, respectively). This is also traceable in a higher value of the spectral hardness: $HR = 2.2\pm0.3$ and $1.7\pm0.3$ for the first and second structures, respectively.

\begin{figure}[h]
	%\vspace{6cm}
	\centering
	\includegraphics[width=10cm]{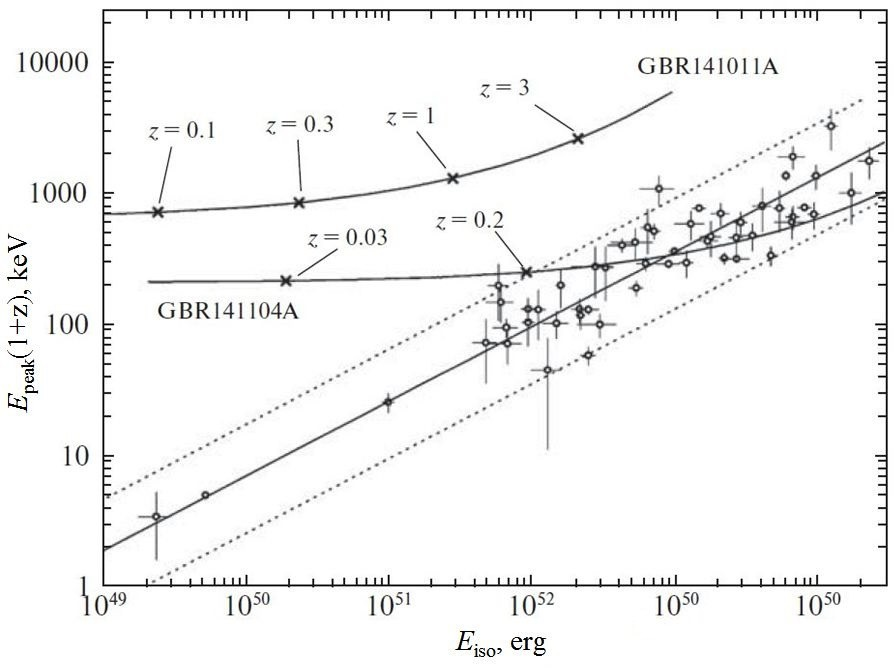} 
	\caption{Amatis diagram, i.e., the dependence of the equivalent isotropic energy Eiso emitted in the gamma-ray range on $E_{peak}(1 + z)$ in the sources reference frame (Amati 2010). The diagram was constructed for long bursts	from the Konus-Wind experimental data given in Volnova et al. (2014). The solid straight line indicates a power-law fit to the dependence; the dotted lines bound the $2\sigma$ correlation region. The trajectories of GRB 141011 and GRB 141104A are plotted as a function of the presumed redshift z.} 
	\label{fig6}
\end{figure}

One of the interesting phenomenological relations for GRBs is Amatis diagram, i.e., the dependence of the equivalent isotropic energy $E_{iso}$ emitted in the gamma-ray range on parameter $E_p(1 + z)$ in the sources reference frame (Amati 2010), where $z$ is the cosmological redshift. Long bursts obey well this law, while short bursts usually lie on the diagram above the main correlation region of long bursts (for short bursts $E_p(1 + z)$ is considerably higher at the same value of Eiso (see, e.g., Amati 2010). Thus, Amati diagram can also be used for the classification of bursts. The redshift of the GRB 141011 source is unknown, but using the fluence estimates in the range 1--10 000 keV and $E_p(1 + z)$, we can construct the trajectory of GRB 141011 on the diagram as a function of z (see Fig. \ref{fig6}). To construct the trajectory, we used the fluence in the range 1--10 000 keV obtained from the GBM/Fermi experimental data with the RMfit v4.3.2 software package, $F=(1.2\pm0.1) 10^{-6}$ erg cm$^{-2}$. We see that the trajectory does not cross the correlation region atany z. Furthermore, the trajectory passes higher, which confirms the classification of GRB 141011 as a short GRB (see also Fig. 1 from Amati (2010)). Since the burst does not fall into the $E_p(1 + z)—E_{iso}$ correlation region at any $z$, the redshift and $E_iso$ cannot be estimated. However, we can give an upper limit for $E_iso$ by assuming that the burst redshift does not exceed $z = 1$ (the median redshift $z$ for a sample of short bursts with reliably determined redshifts was estimated in Berger (2014) and is $< z >= 0.48$). In this case, $E_{iso}(z = 1) = 2.8\cdot10^{51}$ erg. The diagram was constructed from the KONUS-Wind experimental data and is given in Volnova et al. (2014). Here and below, we used the standard cosmological model $[\Omega_\Lambda, \Omega_M , h] = [0.714, 0.286, 0.696]$.

The detectors of the DRGE-1 instrument recorded this GRB on November 4, 2014, the onset of a rise in flux and its maximum were observed at $\sim$00 : 03 : 20 and $\sim$00 : 03 : 29 (UT), respectively. This time is consistent with the burst detection times in the BAT/Swift experiments and other ex- periments presented in GCN circulars (\verb|https://gcn.gsfc.nasa.gov/other/141104A.gcn3|). The slight deviation from the KONUS-Wind detection time (the trigger at 00 : 03 : 19.517 UT) is associated with the trajectory of the Wind spacecraft, which was at a distance of several light-seconds from the Earth at the instant of burst detection. This burst, along with GRB 141011, was seen at a large angle to the detector axis, which was $83^{\circ}$ from the direction toward the nadir. The sources coordinates were used to determine this angle: $R.A. =$ 18 h 37 m 57 s and $Dec. = -12^{\circ} 42^{\prime} 07^{\prime\prime}$, which were obtained by triangulation from the data of several spacecraft (Hurley et al. 2014).

\begin{figure}
	%\vspace{6cm}
	\centering
	\includegraphics[width=8cm]{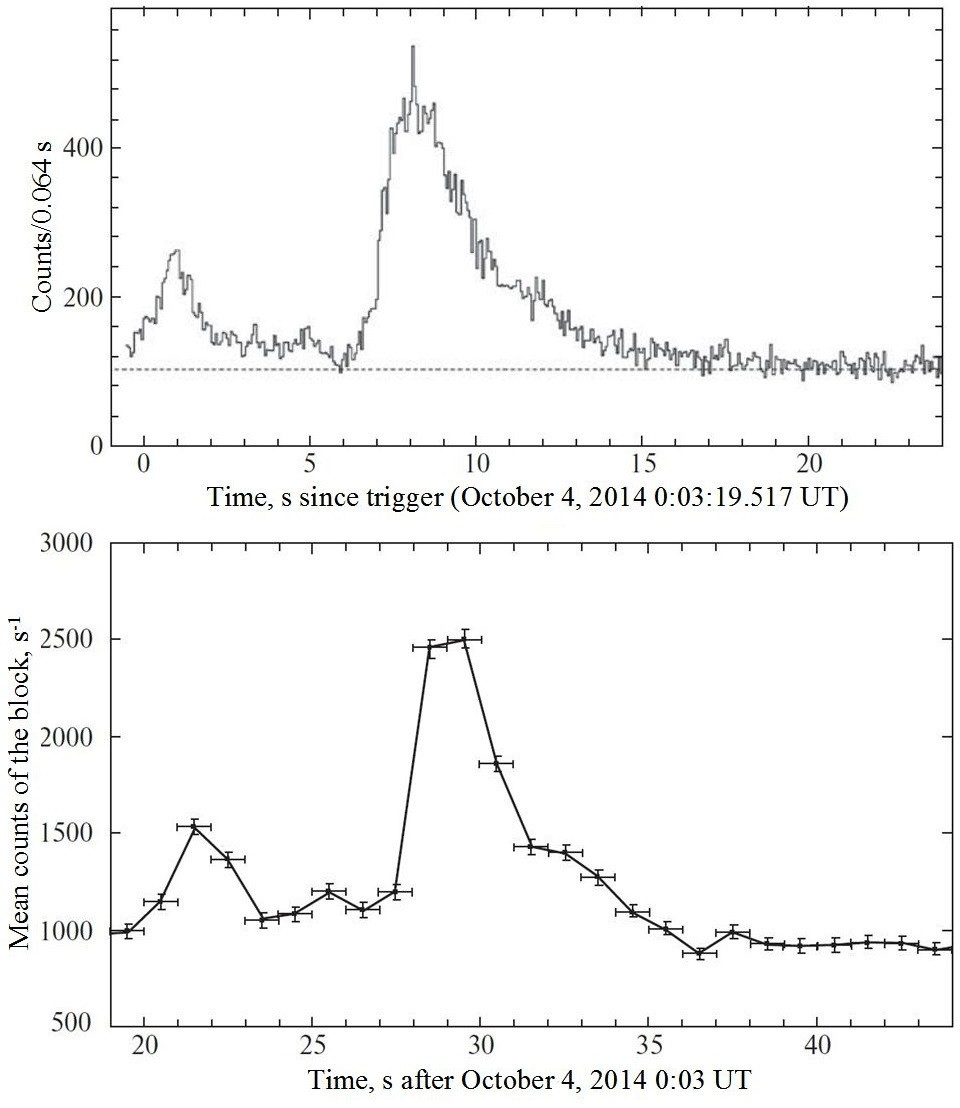} 
	\caption{ Light curves of GRB 141104A obtained onboard the Vernov satellite (the monitoring channels of the	DRGE-11 and DRGE-12 detectors of the RELEC instrumentation, lower panel) and KONUSWind (18--1160 keV,http://www.ioffe.ru/LEA/GRBs/GRB141104\_T00199) (upper panel).} 
	\label{fig7}
%\end{figure}
%\begin{figure}[h]
	%\vspace{6cm}
	\centering
	\includegraphics[width=10cm]{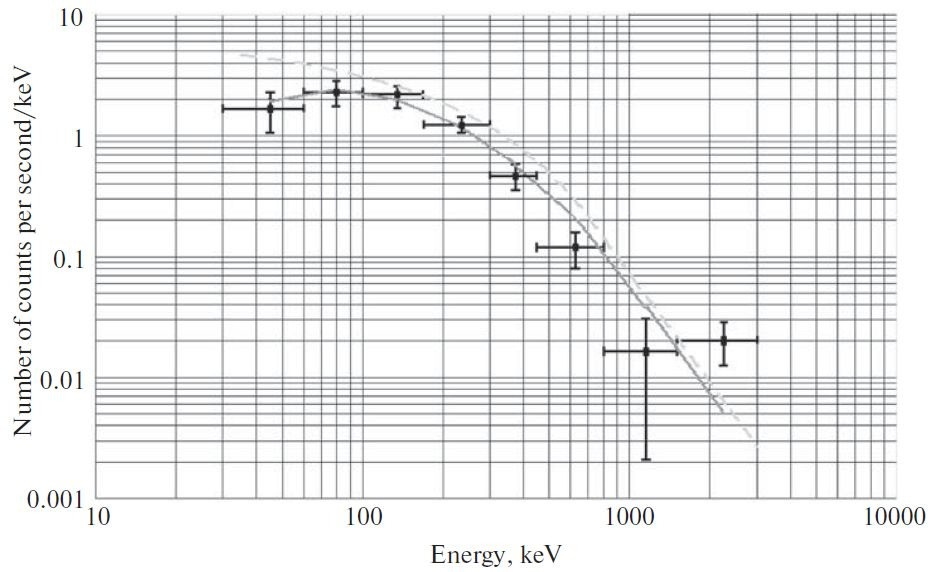} 
	\caption{ Energy spectrum of GRB 141104A based on the Vernov satellite data. The intervals indicate the experimental data: the count rate in Cs I(Tl). The dashed line is the GRB spectrum (BAND model). The solid line is a convolution of the model function with the instrument response matrix.} 
	\label{fig8}
\end{figure}

The burst time profile presented as the integrated count rate of various scintillators of the DRGE-1(2) detector units and constructed from the monitoring data with a time resolution of 1 s is shown in Fig. \ref{fig7} (lower panel). Note that two peaks with a total duration of $\sim$25 s are identified in the light curve of GRB 141104A. Such structures are also seen in the Konus-Wind time profile presented in Fig. \ref{fig7} (upper panel) (http://www.ioffe.ru/LEA/GRBs/GRB141104 T00199). The small peak approximately at 0 : 03 : 25 UT is also significant, as are the local maxima at $\sim$00 : 03 : 32 and, probably, $\sim$00 : 03 : 37 UT. These peaks cannot be associated with magnetospheric phenomena, because they are also seen in the KONUS-Wind burst profile.
The burst energy spectrum constructed for the burst peak (00 : 03 : 28–00 : 03 : 30 UT) is shown in Fig. \ref{fig8}, which also presents its fit by the BAND model. The model parameters are: $E_p=200\pm60$ keV, $\alpha=-0.2\pm0.1$, and $\beta=-2.4\pm0.4$. According to the KONUS-Wind data, the spectrum of this GRB in the energy range 20 keV--15 MeV is well fitted by a BAND model with parameters $\alpha=-1.09(-0.11/+ 0.13), \beta=-2.57(-0.25/+ 0.15)$ and $E_p=160(-16/+ 13)$ keV (Golenetskii et al. 2014c). The discrepancies in the spectral indices for low energies (alpha) obtained in the experiments onboard the Vernov and KonusWind satellites may be associated with an insufficient accuracy of the fit at low energies, because for this burst it turned out to be possible to use only the events in the Cs(Tl) crystals, which detect photons in a harder energy range than do the NaI(Tl) detectors of the KONUSWind experiment. The parameters obtained when the spectrum of GRB 141104A was fitted by other models are also given in Table \ref{GRB}. Just as for GRB 141011A, the fit by a power law (PL) turned out to be considerably poorer than the fits by a power law with an exponential cutoff (CPL) and a BAND model.

The redshift of GRB 141104 was not measured. Using the estimates of $E_{iso}(z)$ and $E_p(z)$ obtained from the DRGE data ($E_p$ and the fluence in the range 20--3000 keV are given in Table \ref{GRB}), we can construct the trajectory of GRB 141104 on Amatis diagram as a function of z similar to that constructed previously for GRB 141011A (Fig. \ref{fig6}). The trajectory is seen to cross the correlation region at $z\sim0.2$, i.e., at $z > 0.2$ GRB 141104A is in the $2\sigma$ region of Amatis diagram for long bursts. Thus, for GRB 141104 we can estimate an lower limit for the redshift $z = 0.2$ and the parameter $E_{iso}(z = 0.2) = 9\cdot10^{51}$erg.

%*************************************************************
\section{DISCUSSION AND CONCLUSIONS}
\noindent
Two GRBs, GRB 141011A and GRB 141104A, confirmed by other space observatories were de- tected during the DRGE experiment onboard the Vernov satellite from July to December 2014. Our processing of the data for these events and their comparison with the GBM/Fermi and KONUS- Wind experimental data show good agreement between the burst parameters. The data obtained allowed us to determine the physical parameters of the bursts, namely the limits of the equivalent isotropic energy for the burst and the redshift of the GRB 141104A source ($z > 0.2$) and the minimum variability time scale (2 ms) of the short GRB 141011A.
Despite the fact that because of the orbital parameters of the Vernov satellite, the active operation time of the DRGE experiment is shorter than, for example, the GBM/Fermi operation time due to a shorter useful time and a larger background, this orbit allows the experiments aimed at detecting cosmic GRBs to be conducted. This turned out to be an important experience for the DRGE experiment onboard the Lomonosov satellite placed in a similar orbit.
Continuous coverage of the celestial sphere is needed to detect short GRBs, whose sources, as is believed, can also be the sources of gravitational-wave events associated with the mergers of compact relativistic neutron star-neutron star or neutron star-black hole binary systems. This cannot be provided by a circumterrestrial experiment alone (for example, GBM/Fermi) but can be provided by a system of spacecraft. The detection and monitoring of short bursts are a significant contribution of the Vernov and Lomonosov satellites for the task of searching for the electromag- netic component of gravitational-wave events predicted for detection already in the immediate future by the LIGO and Virgo facilities (Abbott et al. 2016).

\section*{ACKNOWLEDGMENTS}
\noindent
The RELEC experiment onboard the Vernov satellite was conducted within the framework of the Federal Space Program of the Russian Federation and funded by the Federal Space Agency. We thank the Space Monitoring Center of the Skobeltsyn Institute of Nuclear Physics of the Moscow State University for the creation of the RELEC database and D. Frederiks for useful discussions. The work of P.Yu. Minaev was supported by the Russian Foundation for Basic Research (grant no. 16-32-00489 mol\_a).

\pagebreak   



\begin{thebibliography}{00}
	
	\bibitem{1} B. P. Abbott, R. Abbott, T. D. Abbott, M. R. Abernathy, F. Acernese, K. Ackley, C. Adams, T. Adams, et al., arXiv:1607.07456 [astro-ph.HE] (2016).
	
	\bibitem{2} L. Amati, arXiv:1002.2232 [astro-ph.HE] (2010).
	
	\bibitem {3} D. L. Band, Astrophys. J. 486, 928 (1997).
	
	\bibitem {4} D. Band, J. Matteson, L. Ford, B. Schaefer, D. Palmer, B. Teegarden, T. Cline, M. Briggs, et al., Astrophys. J. 413, 281 (1993).
	
	\bibitem {5} E. Berger, Ann. Rev. Astron. Astrophys. 52, 43 (2014).
	
	\bibitem {6} V. V. Bogomolov, M. I. Panasyuk, S. I. Svertilov, et al., Kosm. Issled. (2017, in press).

	\bibitem {7} W. Cash, Astrophys. J. 228, 939 (1979).	

	\bibitem {8} L. A. Ford, D. L. Band, J. L.Matteson, M. S. Briggs, G. N. Pendleton, R. D. Preece, W. S. Paciesas, B. J. Teegarden, et al., Astrophys. J. 439, 307 (1995).
	
	\bibitem {9} S. Golenetskii, R. Aptekar, V. Palshin, D. Frederiks, D. Svinkin, T. Cline, K. Hurley, J. Goldsten, et al., GCN Circ., No. 16906 (2014a).
	
	\bibitem {10} S. Golenetskii, R. Aptekar, D. Frederiks, V. Palshin, P. Oleynik, M. Ulanov, D. Svinkin, A. Tsvetkova, et al., GCN Circ., No. 16907 (2014b).
		
	\bibitem {11} S. Golenetskii, R. Aptekar, D. Frederiks, V. Palshin, P. Oleynik, M. Ulanov, D. Svinkin, A. Tsvetkova, et al., GCN Circ., No. 17027 (2014v).
	
	\bibitem {12} J. Hakkila and R. Preece, Astrophys. J. 740, 104 (2011).
	
	\bibitem {13} K. Hurley, J. Goldsten, S. Golenetskii, R. Aptekar, V. Palshin, D. Frederiks, D. Svinkin, T. Cline, et al., GCN Circ., No. 17026 (2014).
	
	\bibitem {14} V. V. Khartov, Vestn. NPO im. S. A. Lavochkina, No. 3.3 (2011).
	
	\bibitem {15} A. von Kienlin, C. A. Meegan, W. S. Paciesas, P. N. Bhat, E. Bissaldi,M. S. Briggs, J. M. Burgess, D. Byrne, et al., Astrophys. J. Suppl. Ser. 211, 13 (2014).
	
	\bibitem {16} A. von Kienlin, GCN Circ., No. 16905 (2014).
	
	\bibitem {17} P. Minaev, A. Pozanenko, and V. Loznikov, Astron. Lett. 36, 707 (2010).
	
	\bibitem {18} P. Minaev, A. Pozanenko, S. Molkov, and S. Grebenev, Astron. Lett. 40, 235 (2014).
	
	\bibitem {19} P. Minaev and A Pozanenko, Astron. Lett. 43, 1 (2017).
	
	\bibitem {20} B. Paczynski, Astrophys. Lett. 494, 45 (1998).
	
	\bibitem {21} M. I. Panasyuk, S. I. Svertilov, V. V. Vogomolov, G. K. Garipov, E. A. Balan, V. O. Barinova, A. V. Bogomolov, I. A. Golovanov, et al., Adv. Space Res. 57, 835 (2016).
	
	\bibitem {22} D. S. Svinkin, D. D. Frederiks, R. L. Aptekar, S. V. Golenetskii, V. D. Palshin, Ph. P. Oleynik, A. E. Tsvetkova, M. V. Ulanov, et al., arXiv:1603.06832 [astro-ph.HE] (2016).
	
	\bibitem {23} E. Troja, S. Rosswog, and N. Gehrels, Astrophys. J. 723, 1711 (2010).
	
	\bibitem {24} A. A. Volnova, A. S. Pozanenko, J. Gorosabel, D. A. Perley, D. D. Frederiks, D. A. Kann, V. V. Rumyantsev,
	V. V. Biryukov, et al., Mon. Not. R. Astron. Soc. 442, 2586 (2014).
	
	\bibitem {25} S.Woosley, Astrophys. J. 405, 273 (1993).
	
	\bibitem {26} A. Zoglauer, PhD Thesis (Tech. Univ., Munich, 2005).
	
	\bibitem {27} A. Zoglauer, R. Andritschke, and F. Schopper, New Astron. Rev. 50, 629 (2006).
		
	
\end{thebibliography}


\end{document}